\documentclass[useAMS,usenatbib]{mn2e} 
\pdfoutput=1 
\usepackage{journals}
\usepackage[pdftex]{graphicx,color} 
\usepackage[latin1]{inputenc}
\usepackage{graphics} 
\usepackage{amsfonts} 
\usepackage{amsmath}
\usepackage{multicol} 
\usepackage{layout} 
\usepackage{amssymb}
\title[On the cluster concentration-mass relation]
{Reconciling extremely different concentration-mass relations}
\author[Meneghetti M. \& Rasia E.]
{\parbox{\textwidth}{Massimo Meneghetti$^{1,2}$\thanks{E-mail:
{massimo.meneghetti@oabo.inaf.it}}, Elena Rasia$^{3}$} \\ \\ 
$^{1}$ INAF - Osservatorio Astronomico di Bologna, via Ranzani 1, 40127, Bologna,
 Italy \\ 
$^{2}$ INFN - Sezione di Bologna, viale Berti Pichat 6/2, 40127,
 Bologna, Italy \\
$^3$ Department of Physics, University of Michigan, 450 Church St., Ann Arbor, MI  48109 
 }
\begin{document}
\date{}
\maketitle
\label{firstpage}
\pagerange{\pageref{firstpage}--\pageref{lastpage}} \pubyear{2012}
\begin{abstract}
The concentration-mass relations proposed by Prada et al. (2012) and by Duffy et al. (2008) on the scales of galaxy clusters show  some of the largest discrepancies among all the works present in literature. This is surprising because they are both derived from the analysis of dark-matter halos forming in $\Lambda$CDM simulations with similar set-ups. 
With the help of  analytic models and numerical simulations we investigate the origin of this discrepancy focusing on the procedures used to derive the concentrations (circular velocity ratios versus density profile fitting) and on the selection criteria used to bin the halos (binning in maximum circular velocity versus binning in mass).
We find that both steps of the analysis have a large impact on the resulting $c-M$ relation.
In particular, we show that the two $c-M$ relations can be entirely reconciled, if we account for these methodological differences. Our analysis demonstrates that the  concentration estimates are sensitive to the {largely different} radial scales probed by a particular measurement method. This implies that concentrations derived with different techniques (both in observations and in simulations) must be compared over the same radial range.
\end{abstract}


\section{Introduction}

About fifteen years ago \cite{NA96.1} showed that dark-matter halos forming in simulations of hierarchically clustering universes develop a nearly universal density profile, reasonably well described by the formula
\begin{equation}
	\rho(R)=\frac{\rho_s}{R/R_s(1+R/R_s)} \;,
	\label{eq:nfw}
\end{equation}
where $\rho_s$ and $R_s$ are the halo characteristic density and  scaling radius, respectively. This result holds over a wide range of masses, from the scale of massive galaxies to that of galaxy clusters. \cite{NA96.1} also found that low mass halos are centrally denser than their high-mass counter-parts. In particular, the halo {\em concentration}, defined as $c_{200}=R_{200}/R_s$, was found to be a decreasing function of the halo mass, $M_{200}$\footnote{$R_{200}$ and $M_{200}$ are the radius and the mass of the sphere enclosing a mean density equal to 200 times the critical density of the universe.}. 
This outcome was interpreted as the signature that a dark matter (DM) halo keeps memory of the density of the universe at the epoch of its collapse. For this reason, the $c-M$ relation is connected to and provides information on the cosmological model within which halos {form} \citep{NA97.1}.

Virtually, all cosmological observations of the universe on large scales undertaken in the last decade support the so-called concordance cosmological model ($\Lambda$CDM). Simulated universes reproducing variations of this model have been performed to calibrate the $c-M$ relation \citep{NA97.1,BU01.1,EK01.1,zhao2003,2008MNRAS.390L..64D,2008MNRAS.387..536G,2008MNRAS.391.1940M,2011arXiv1112.5479B,2011ApJ...740..102K}. Most theoretical works show relatively similar behaviors with the exception of the recent analysis of  \cite{2012MNRAS.tmp.3206P}, which differs form the others both in terms of amplitude and shape of the $c-M$ relation, especially on the scales of galaxy clusters.  

In this paper, we investigate the origin of this discrepancy focusing on the comparison between the relations of \citet[][P12]{2012MNRAS.tmp.3206P} and \citet[][D08]{2008MNRAS.390L..64D}. 
Both works are based on the analysis of a large number of DM halos extracted from cosmological boxes. For the scope of studying clusters, P12 recurred to the {\tt Multidark} \footnote{http://www.multidark.org} simulation
consisting of a $1 h^{-1}$ Gpc$^3$ box filled with $2048^3$ DM particles. Instead,
D08 used  smaller boxes, the largest one having a side-length of $400 h^{-1}$Mpc filled with $512^3$ particles. {These simulations are therefore characterized by a worse resolution and smaller statistical power on the cluster scales.} 
The simulation codes have different architectures (mesh-based, {\tt ART} - \citealt{1997ApJS..111...73K}, in the first case and particle-based, $\tt GADGET$ - \citealt{SP01.1}, in the second). However, since a long time ago, it has been demonstrated that the DM properties are unaffected by the numerical scheme \citep{frenk99}.
Finally, the differences in terms of cosmological parameters are small\footnote{$(\Omega_m,\Omega_b,\Omega_{\Lambda},h,\sigma_8,n_s)$ equal to (0.27, 0.046, 0.73, 0.70, 0.82, 0.95) for P12 and (0.26, 0.046, 0.74, 0.80, 0.80, 096) for D08}. 

That said, it is surprising that the P12 concentrations are $\sim50-60\%$ higher than the  concentrations found by D08 at fixed halo mass. The P12 relation, additionally, exhibits an upturn towards the largest masses: after a certain ``pivot'' mass the logarithmic slope of the relation from negative becomes positive. Such behavior is particularly evident at high redshift, as the pivot mass decreases with redshift.

Some discussion on the impact of selection effects on the shape of the P12 $c-M$ relation can be found in other works \citep[e.g.][]{2012arXiv1206.1049L}. Here, we aim at fully shedding light on the reasons of the mismatches, discussing the differences between the two methods employed by P12 and by D08 to measure concentrations and to construct the $c-M$ relation (Sect.~\ref{sect:diff}).  
{To be immune from numerical nuisances,} in Sect.~\ref{sect:analysis}, we study the systematics introduced by the two procedures utilizing both analytic models and {the halos  from the best resolved  {\tt Multidark} simulation}.  Finally, in Sect.~\ref{sect:conclu} we summarize the results and draw our conclusions.

\section{Procedural differences}
\label{sect:diff}



\subsection{Methods to derive the concentration} 

\label{sect:pradameth}

{\bf{Prada et al. (2012)}}'s procedure is described by the following steps:\\
\noindent {\bf 1)} each halo is identified in the cosmological box and characterized  by means of its $R_{200}$ radius and $M_{200}$ mass; \\ 
\noindent {\bf 2)} two velocities are computed from the mass profile in logarithmically equi-spaced radial bins:  

$i)$ the halo maximum circular velocity: $V_{max}=\max (V_{circ}$), where $V_{circ}=\sqrt{{GM(<R)}/{R}}$, and
  
$ii)$ the circular velocity at $R_{200}$:
		$V_{200}=\sqrt{{GM_{200}}/{R_{200}}}$. 
Assuming a NFW profile, the two velocities are related through the concentration in a way that more centrally concentrated halos are characterized by higher $V_{max}/V_{200}$ ratios: 
	\begin{equation}
		\frac{V_{max}}{V_{200}}=\sqrt{\frac{0.216c}{f(c)}}=F(c) \;,
			\label{vmaxc}
	\end{equation}		
	where $f(c)$ is given by
		$f(c)=\ln(1+c)-{c}/(1+c)$; \\
\noindent {\bf 3)} finally, the halo concentration is derived by numerically inverting Eq.~\ref{vmaxc}. P12 concentration ($c_{P12}$) is, therefore, estimated from {\it{mass measurements done only at $R_{200}$ and at $R_{max}$}}. \\

%



\noindent{\bf{Duffy et al. 2008}}, instead, followed a more standard methodology. They searched for the NFW best-fit of the halo density profiles (Eq.~1) evaluated in equally logarithmically-spaced radial bins, with  $-1.25 \leq \log{R/R_{vir}} \leq 1$. 
The concentration ($c_{D08}$) is one of free parameters of the fit (the other is the normalization $\rho_s$, Eq.~\ref{eq:nfw}). 


\subsection{Binning}
\label{sect:pradarel}

On top of the concentration derivation, the procedures of P12 and D08 differ also for the selection criteria adopted to bin the halos. 
In the former case, halos are sorted by $V_{max}$ ($c_{V sel}$), while in the latter via mass ($c_{M sel}$).

This has important consequences.
Using Eq.~\ref{vmaxc} and the circular velocity definition, we can see  that 
      \begin{equation}
		V_{max} =  V_{200}\sqrt{\frac{0.216c}{f(c)}}=\sqrt{\frac{GM_{200}}{R_{200}}}F(c) \propto M_{200}^{1/3}F(c) \;,
     \end{equation} 
Thus, binning in $V_{max}$ translates into selecting curved regions in the concentration-mass plane as shown in Fig.~1 where some curves at constant $V_{max}$ (from $1000$ km/s to 1500 km/s) are plotted.   
Note that the curves are steeper at the high-mass end while they flatten towards the smaller masses. Consequently, 
we expect  the median concentrations in $V_{max}$ bins to be larger than those measured in mass bins, because the same $V_{max}$ bin also contains many systems having slightly smaller mass but much higher concentrations. 

\section{Understanding the differences}
\label{sect:analysis}

\subsection{Consequences of the binning}

To illustrate the effect of  the binning method, 
we use analytic prescriptions and Monte-Carlo simulations to generate 
halos drawing them from the Sheth \& Tormen mass function \citep{SH02.1} and associating them 
the fiducial concentration $\hat{c}(M,z)$ of D08 for relaxed clusters:
\begin{equation}
\hat{c_{200}}(M_{200},z)=\frac{6.71}{(1.+z)^{0.44}}\left(\frac{M_{200}}{2\times 10^{12}h^{-1}M_\odot}\right)^{-0.091}
\label{eq:duffy}
\end{equation}
Further, we assume that the concentrations at fixed mass are log-normally distributed,
\begin{equation}
	p(c)=\frac{1}{c\sqrt{2\pi\sigma^2}}\exp\left[-\frac{(\ln{c}-\ln\hat{c})^2}{2\sigma^2} \right]\; ,
\end{equation}
and we change the concentration scatter by varying $\sigma$ between $0.2$ and $0.6$ in steps of 0.1. Numerical simulations are usually consistent with $\sigma\sim 0.25-0.4$ \citep{DO03.2,deboni.etal.12}, almost independently of mass. We find that the {\tt Multidark} halos used by P12 have a concentration scatter of $\sigma\sim0.35$. 
To focus exclusively on the problematics of the binning, we do not measure the concentrations of these halos {\em \`a la } P12, but we adopt their binning method to re-construct the $c-M$ relation. We verified that the choice of another mass function or concentration-mass relation to generate the halos does not change the results of the following analysis.\\

{\bf Halos at low redshift.}
According to previous equations, we begin our analysis by building concentration distributions for different values of $\sigma$ and for halos at redshift $z=0.25$.  First, we select halos in a narrow bin of circular velocity: $1000 \;{\rm km/s} < V_{max} \le 1100 \;{\rm km/s}$. The median mass in the bin varies little as a function of the concentration scatter, being $2.1 \times 10^{14}\;h^{-1}M_\odot$ for $\sigma=0.2$ and $1.9 \times 10^{14}\;h^{-1}M_\odot$ for $\sigma=0.6$. Using Eq.~\ref{eq:duffy}, these mass values correspond to D08 concentrations of $\hat{c}_{M sel}\sim 3.96$ and $\hat{c}_{M sel}\sim 3.99$ respectively, {\it but} due to the positive skewness of the concentration distributions in the $V_{max}$ bin, we measure median concentrations for the selected halos growing from $c_{D08,V sel}=$ 4.0 for $\sigma=0.2$ to $c_{D08,V,sel}=$5.2 for $\sigma=0.6$.  

These results confirm that  median concentrations are larger in  $V_{max}$-selected  than in mass selected samples of halos, even if the median mass in the bin remains (almost) constant. Moreover, the ratio between median concentrations in $V_{max}$- and mass-selected samples is a growing function of the concentration scatter of the parent halo population. 

\begin{figure*}
\begin{center}
\includegraphics[width=0.33\hsize]{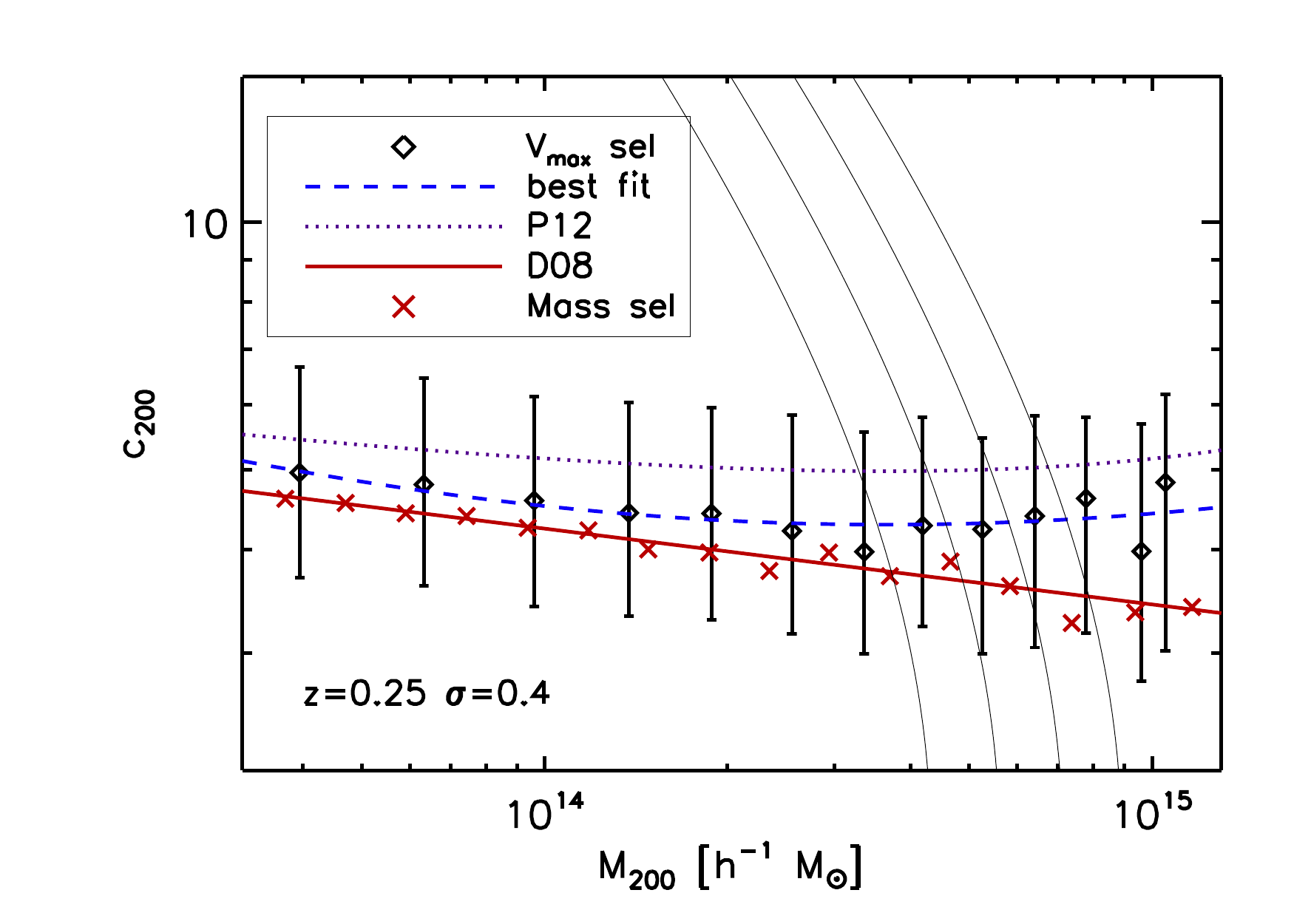}
\includegraphics[width=0.33\hsize]{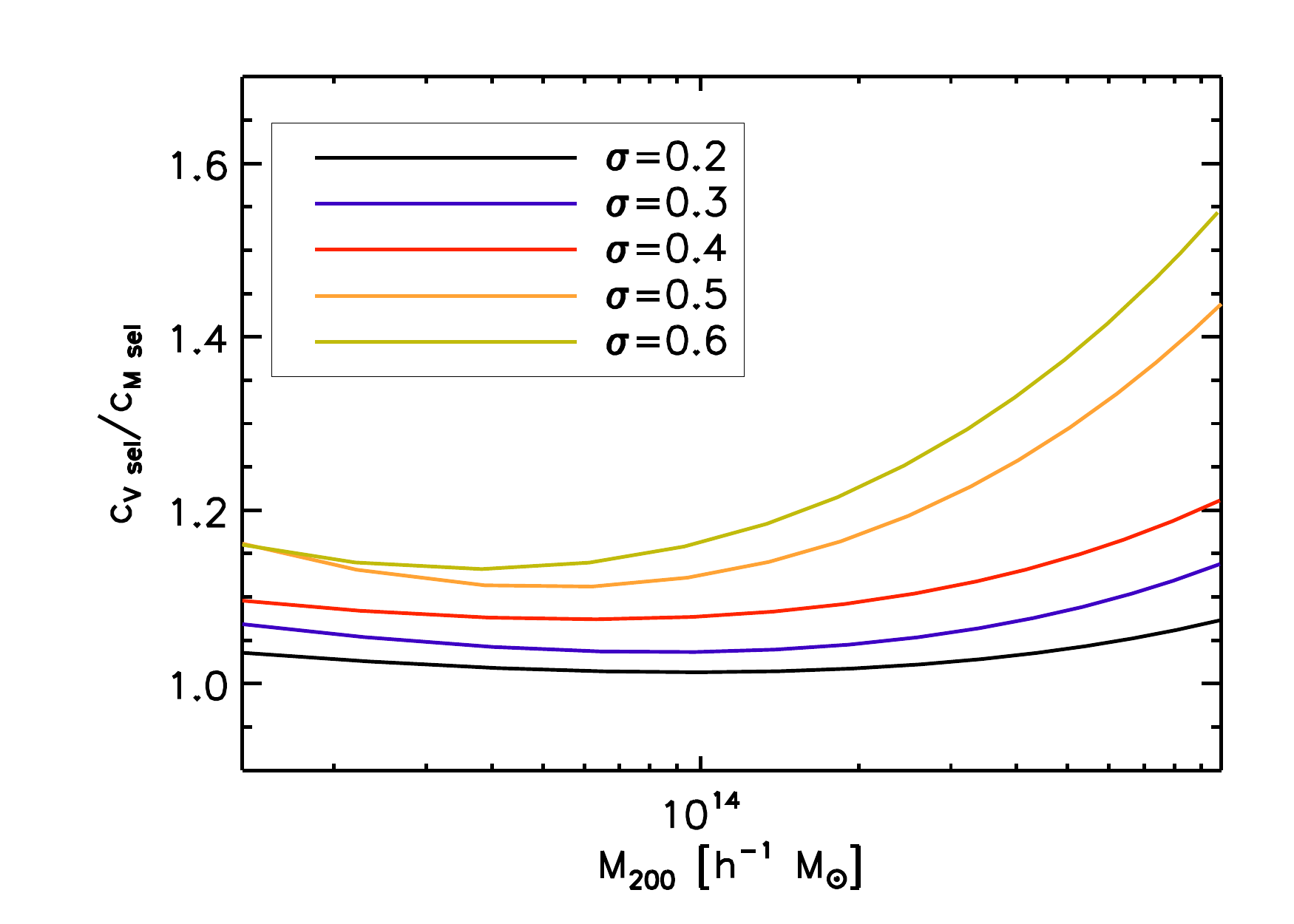}
\includegraphics[width=0.33\hsize]{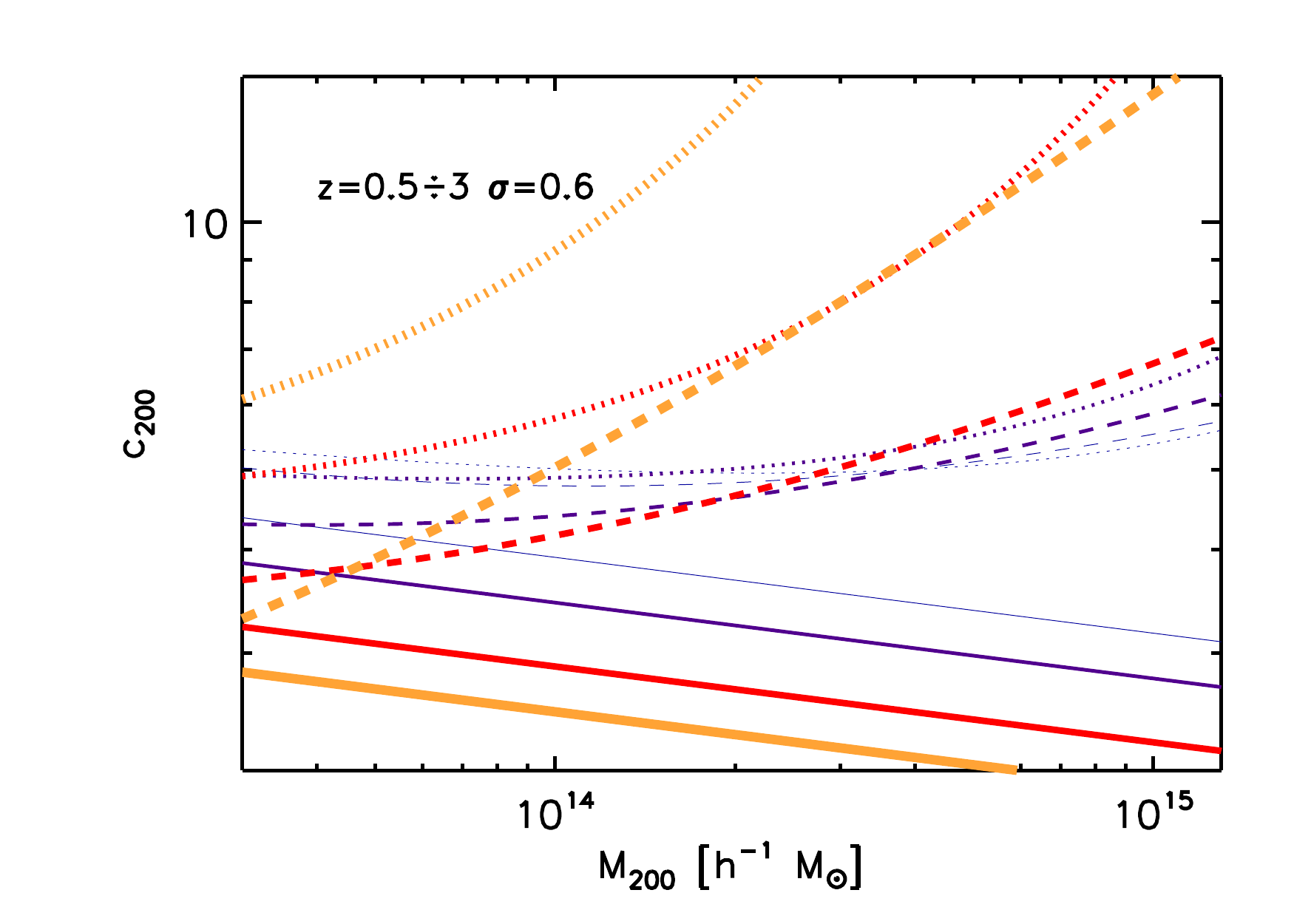}
\caption{Left panel: The $c-M$ relation of $V_{max}$ (diamonds) and mass-selected halos (crosses) at $z=0.25$. Results are shown for $\sigma=0.4$. The solid-red and the dotted-blue lines represent the D08 and P12 $c-M$ relations, respectively. The dashed-blue line indicates the best-fit to the diamond data points. We also show some curves of constant $V_{max}$ (thin solid lines). Central panel: For different values of $\sigma$, ratio between the concentrations of halos selected by either $V_{max}$ or mass as function of mass. Right panel: Evolution of the $c-M$ relation of $V_{max}$-selected halos as a function of redshift (dashed lines). Results are shown for $z=0.5,1,2,3$ and refer to the case $\sigma=0.6$. Ticker lines (red-yellowish colors) correspond to higher redshifts. The solid and the dotted lines show the redshift evolution of  the D08 and of the P12 $c-M$ relations.}
\label{fig:vmaxcm}
\end{center}
\end{figure*}

In the left panel of Fig.~\ref{fig:vmaxcm}, we generalize these results for the entire population of halos assuming a concentration scatter $\sigma=0.4$.
The medians of the concentrations in each $V_{max}$ bin are shown as black diamonds with the associated  inter-quartile ranges (vertical bars).
The blue dashed line represents the best-fit to the data using the P12's functional: 
\begin{equation}
c_{200}=\alpha\left(\frac{M_{200}}{10^{12}h^{-1}M_\odot}\right)^{\beta}\left[1+\gamma\left(\frac{M_{200}}{10^{15}h^{-1}M_\odot}\right)^{1/2}\right]\;,
\label{eq:p12func}
\end{equation}
The original relations by P12 and D08 are shown by the dotted-magenta and by the solid-red lines respectively. Finally, as a consistency check, the concentrations derived by binning the sample in mass are plotted with red crosses. As expected these data points perfectly reproduce the input D08 $c-M$ relation.
As discussed above, we find that for any mass the concentrations measured in $V_{max}$ bins are higher than those derived in mass bins. 
The ratio of concentrations in $V_{max}$ ($c_{\rm V sel}$) and in mass bins ($c_{\rm M sel}$) grows with the concentration scatter as shown in the central panel of Fig.~\ref{fig:vmaxcm}. It also depends on the halo mass, being a slowly decreasing function of $M_{200}$ at low masses, but with a clear upturn at the highest masses.  
For $\sigma\sim 0.3-0.4$, typical of numerically simulated halos, the concentrations of the $V_{max}$-selected halos are higher than those of the mass-selected ones by an amount that grows from $5\%$ to $20\%$ between $M=10^{14}\;h^{-1}M_\odot$ and $M=10^{15}\;h^{-1}M_\odot$. At the largest masses, the ratio is strongly dependent on $\sigma$, and it changes from $\sim 5\%$ to $\sim55\%$ as $\sigma$ grows from 0.2 to 0.6.   \\

{\bf Halos at higher redshift.} P12 showed that the upturn in their $c-M$ relation becomes more prominent at high redshifts. We repeat the same procedure to generate halos at redshifts $z=0.5,1,2$ and $3$ (although the D08 relation has been calibrated up to $z=2$). The results in these cases are shown in the right panel of Fig.~\ref{fig:vmaxcm}. To avoid confusion in the figure, we do not show the data points here, but we plot only the best-fit curves, given by the dashed lines, whose thickness increases with redshift. For this analysis, we consider only an extreme concentration scatter, $\sigma=0.6$. We do not claim that this is realistic, but we use it to emphasize the effects of binning. Results for lower values of $\sigma$ can be  extrapolated from the previous discussion. 
  
We clearly see an upturn in the $c-M$ relation when halos are selected by $V_{max}$. The upturn begins at smaller masses as the redshift increases. Interestingly, we see that binning in $V_{max}$ also inverts the redshift evolution of the input D08 $c-M$ relation (solid lines). On the cluster scales, the concentrations of $V_{max}$-selected halos grow with redshift at fixed mass, reproducing the trend which characterizes the P12 $c-M$ relation. For $z=0.5$ and $\sigma=0.6$ binning by $V_{max}$ is sufficient to bring the D08 and the P12 to a full agreement. However, we notice that it becomes harder and harder to reproduce the P12 relation as the redshift increases even assuming an extreme concentration scatter.  
In order to obtain a good match with P12 at $z>0.5$, we would need to modify the initial $c-M$ relation of D08 by increasing its amplitude and  reducing its redshift evolution. We verified that this can achieved by changing the exponent of the $(1+z)$ term in Eq.~\ref{eq:duffy} from 0.44 to 0.25. 

In the literature, there is no consensus on the evolution of the $c-M$ relation with redshift, in particular for massive halos. While several authors predict a strong redshift evolution of the concentrations at all mass scales \citep[see e.g.][]{BU01.1,EK01.1}, \cite{ZH03.1} find that the evolution of the concentration of individual halos is not just a function of redshift but is tightly connected to their mass growth rate \citep[see also][]{2002ApJ...568...52W}. In particular, the faster the mass grows, the slower the concentration increases. Since most of the massive halos are in a fast mass accretion phase at high redshift, they find that the cluster $c-M$ relation has a very slow redshift evolution. Similar results were found recently by \cite{2011MNRAS.411..584M}, who confirm that the growth rate of the concentration depends on the halo mass, with low-mass haloes experiencing a faster concentration evolution. These authors also find that the evolution of the $c-M$ relation is faster at lower redshifts than at higher redshifts.

To summarize this section, we showed that binning in $V_{max}$ instead of binning in mass has two consequences:  
1)  the median concentrations are higher, and 2) there is an upturn at the large mass scales. Both these effects are amplified by increasing the scatter in the concentration distribution. 
Nevertheless, the difference in the binning procedure cannot be the only cause of the discrepancy between P12 and D08. Indeed, as shown in Fig.~\ref{fig:vmaxcm}, for values of $\sigma$ that are typically found in the simulations the increment in the concentration value caused by binning in $V_{max}$ is $\sim15-20\%$, which accounts of about half of the differences between the P12 and D08 $c-M$ relations. 

\subsection{Consequences of the concentration measurement}
Now, we investigate if the remaining discrepancy can be attributed to the different methods employed by P12 and D08 to measure the concentrations (Sect.\ref{sect:pradameth}). To do this, we use numerically simulated halos and, to avoid any extra bias introduced by using different samples, we analyze halos extracted from the Multidark simulation, the same simulation used by P12. 
From their database\footnote{{\tt http://www.multidark.org}}, we query for the masses, concentrations (measured from the $V$-ratio), maximum circular velocities, and mass profiles of all halos at redshift $z=0.25$ with masses satisfying at $M_{200}\geq 10^{14}h^{-1}M_{\odot}$.  

\begin{figure}
\begin{center}
\includegraphics[width=\hsize]{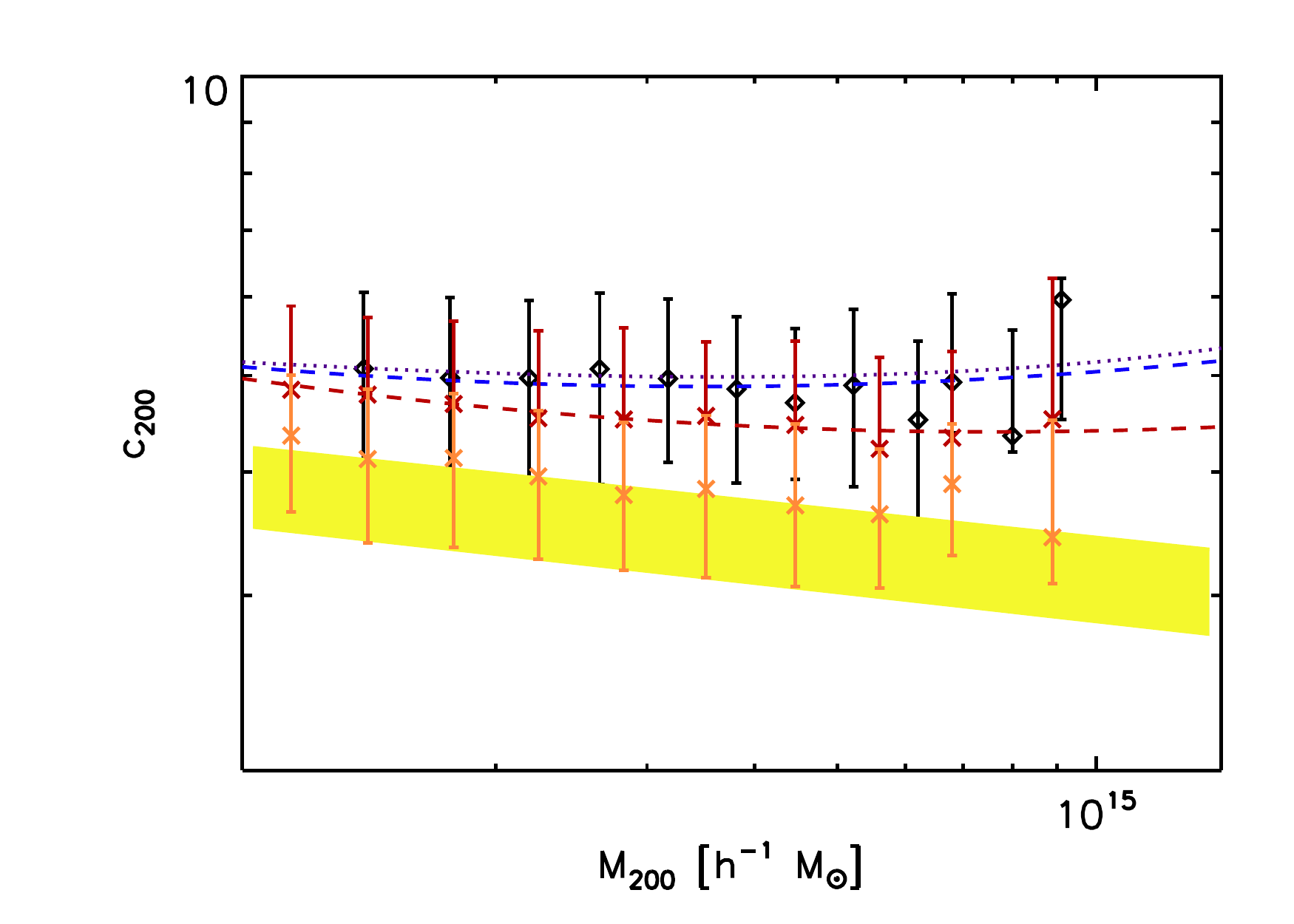}
\caption{The $c-M$ relations of cluster-sized halos extracted from the Multidark simulation at $z=0.25$. The black diamonds show the result of selecting clusters by $V_{max}$. The P12 $c-M$ relation is indicated by the dotted line. Red crosses show the results obtained by binning halos in mass. The best-fit relation is given by the dashed line. Finally, the orange crosses show the median concentrations in mass bins, when the concentrations are measured by fitting the mass profiles with NFW models, i.e. following the approach of D08. The shaded yellow region shows the D08 $c-M$ relation (relaxed and un-relaxed halos).}
\label{fig:multid}
\end{center}
\end{figure}

To corroborate the final concept of the previous section, we create a plot similar to the left panel of Fig.~\ref{fig:vmaxcm}, where the concentration are now derived following the method of P12 (Fig.~\ref{fig:multid}). Black diamonds and red crosses still denote results from the binning in $V_{max}$ and mass, respectively, while dotted-magenta and dashed-red lines report the original relations of P12 and D08.
As consistency check, we can notice that the P12 relation is perfectly recovered. 
Selecting halos by mass, we obtain concentrations lower by $\sim 5-15\%$ compared to P12. 
However, this is still $\sim15-20\%$ higher than the amplitude of the D08 relation. Once again, even if we change the method to derive the concentration, the total discrepancy between D08 and P12 is not explained by simply modifying the binning criterium. As final step, from the mass profile of the Multidark halo we derive the concentration {\it \`a la} D08 and plot the results after binning in mass
(orange crosses). These values are now consistent, within $1\sigma$ with the original D08 relations shown by the yellow band whose upper and bottom limits refer to relaxed clusters and un-relaxed halos. {The fact that they better match the upper limit may be due to the better resolution and larger size of the P12 simulations.}   

The circle is now closed, we demonstrated that the two steps in the procedure to construct the $c-M$ relation are concurring to produce a difference of $\sim 40\%$ between the results of P12 and D08. This result being established, we still need to answer to the question: {\it ``why do different methods to derive the concentration lead to discrepant results?"}

\begin{figure*}
\begin{center}
\includegraphics[width=0.3\hsize]{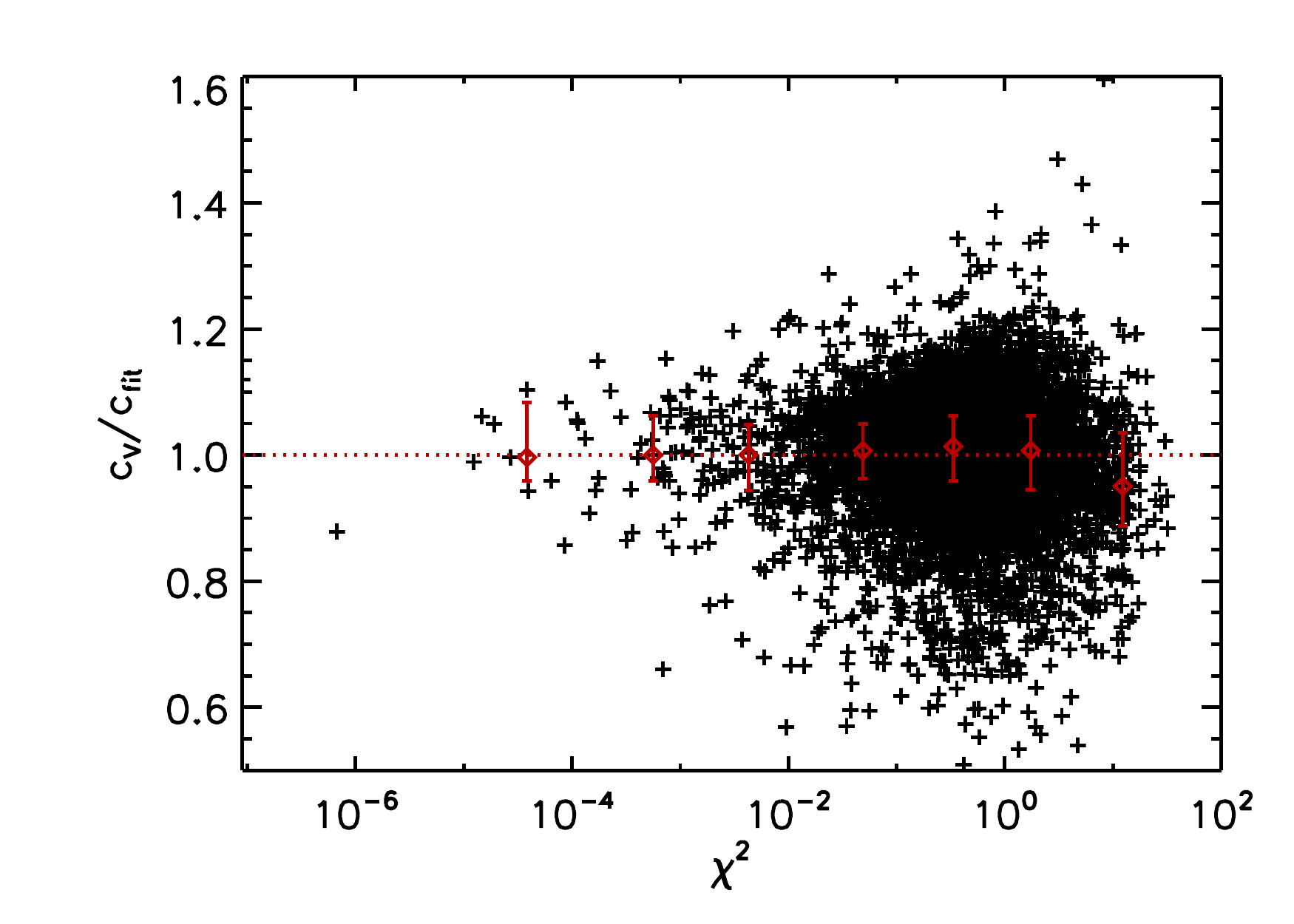}
\includegraphics[width=0.3\hsize]{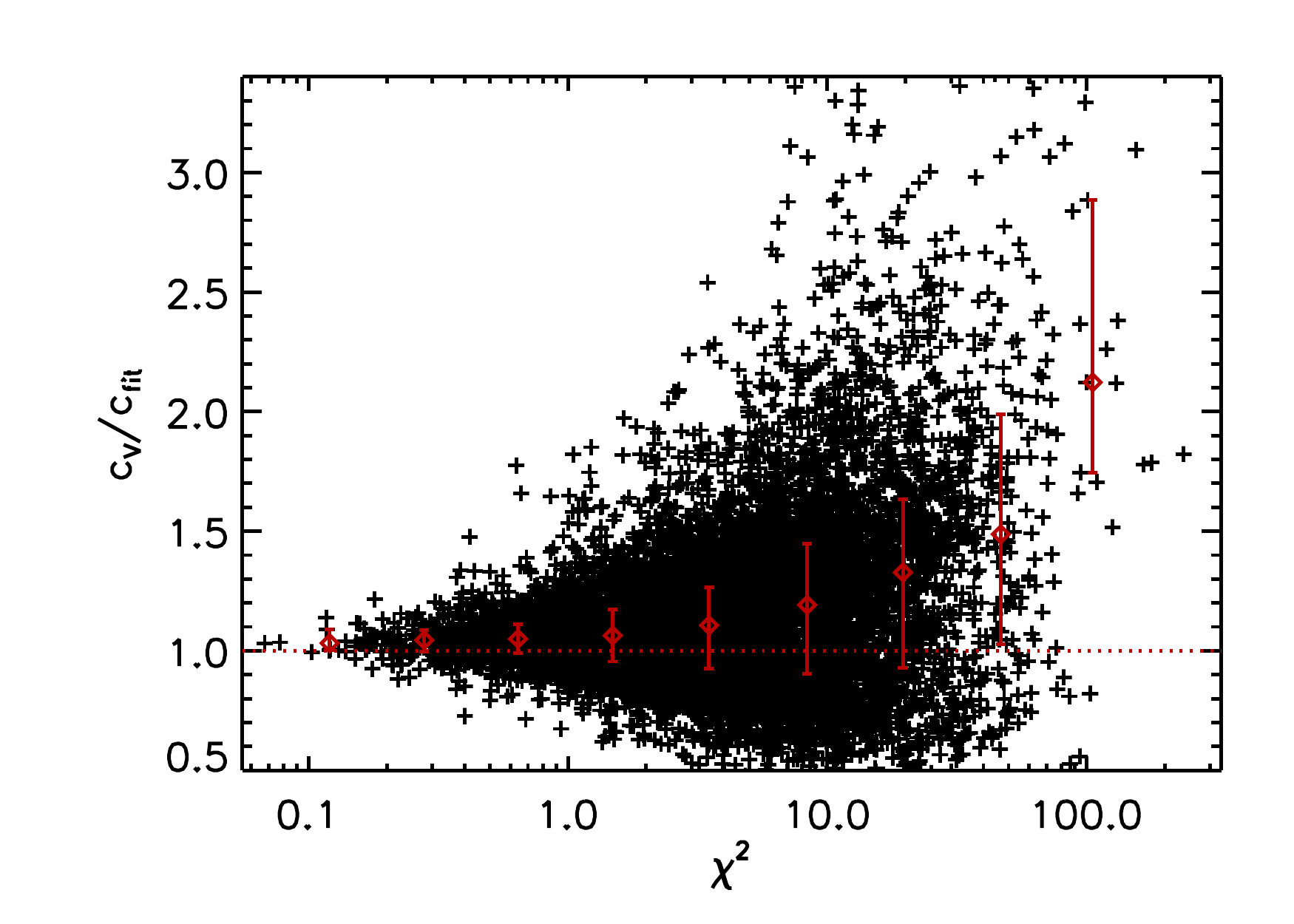}
\includegraphics[width=0.3\hsize]{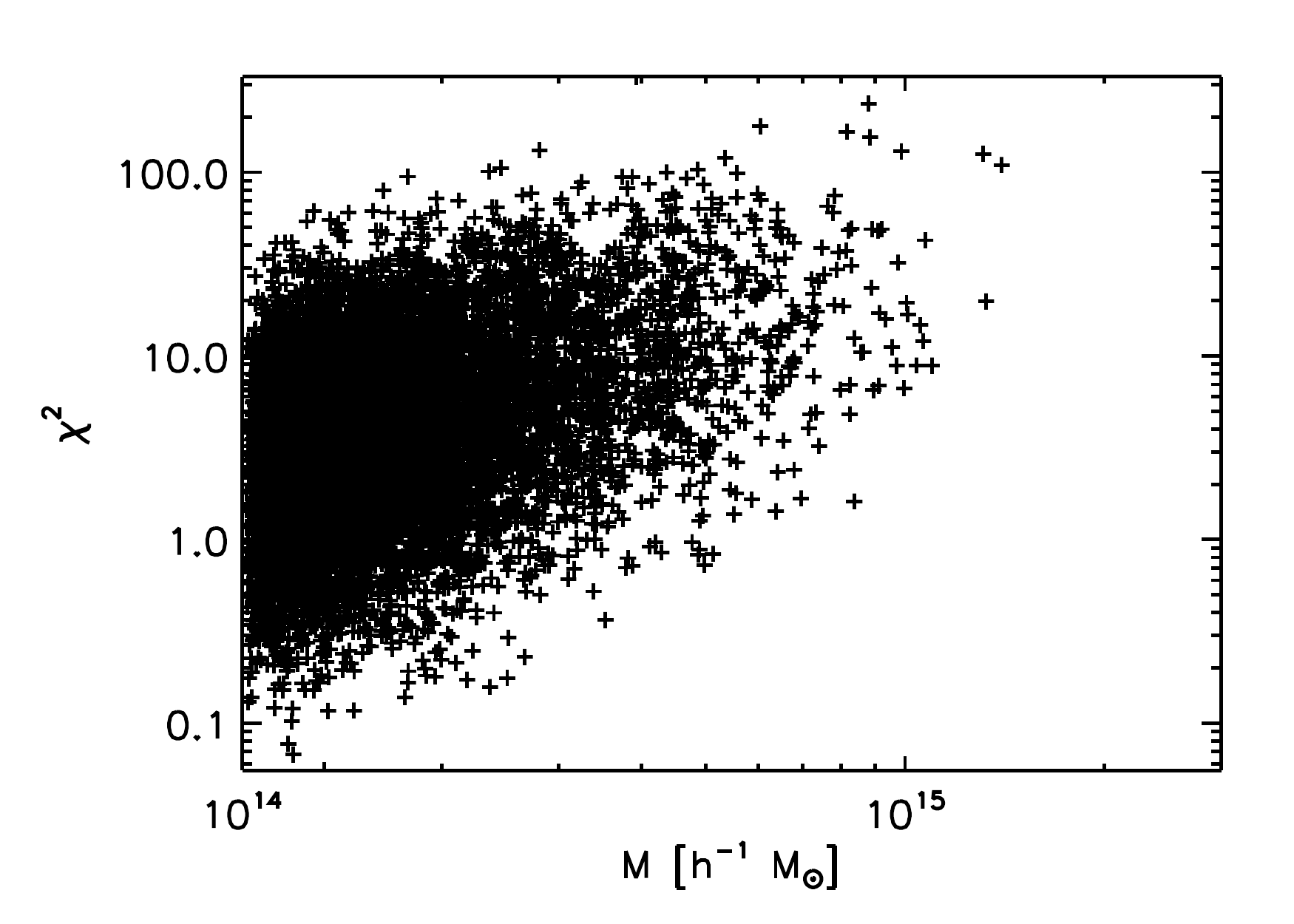}
\caption{Left panel: Ratios between the concentrations measured using the $V$-ratio, as done by P12, and by fitting the halo density profiles with the NFW model in the radial range $[R_{max}-R_{200}]$. The ratios are shown as a function of the $\chi^2$ value of the best fit NFW models. Central panel: as in the left panel, but the fit extends to $0.05R_{vir}$ (as done by D08). Right panel: quality of the best fit as a function of the halo mass.
}
\label{fig:compsim}
\end{center}
\end{figure*}

To discuss this issue we extend the experiment presented in P12 where they compare
the concentrations measured from the $V=V_{max}/V_{200}$ ratio and from the density fit of the 10000 more massive systems in the Bolshoi simulation at $z=0$ and at $z=3$. P12 found that the agreement between the concentrations is good for the highest values of $c$, i.e. for the least massive halos. For $c \lesssim 5$, instead, the concentrations $c_{P12}$ are higher than the $c_{D08}$ by $\sim 10-15\%$. 

Using the sample of halos with $M>10^{14}h^{-1}M_\odot$ at $z=0.25$, we find  that, if we fit the density profiles with the NFW functional in Eq.~\ref{eq:nfw} over the radial range $[R_{max},R_{200}]$, i.e. in the same radial range probed by the concentrations $c_{V}$ of P12, the resulting concentrations $c_{fit}$ are on average in good agreement with those derived from Eq.~\ref{vmaxc}. The results are shown in the left panel of Fig.~\ref{fig:compsim} as function of the $\chi^2$ of the best NFW fit\footnote{The fit is done using the IDL routine {\tt MPFITFUN} -- see {\tt http://cow.physics.wisc.edu/$\sim$craigm/idl/mpfittut.html}}. 

However, we notice that, for an NFW halo, $R_{max}\sim 2R_{s}$ \citep{NA97.1}. Thus, the radial range we just used is significantly different from the radial range within which D08 measured their concentrations. Indeed, they extended the fit to much smaller radii (down to $5\%$ of $R_{vir}$). 
Repeating the fit over the radial range used by D08, we found that {\it only} the halos that are better fitted by an NFW model (i.e. having the smallest $\chi^2$) have $c_{\rm fit}\sim c_{\rm V}$ (central panel of Fig.~\ref{fig:compsim}). This is not surprising, given that Eq.~\ref{vmaxc} holds only for NFW halos. 
On the contrary, there is a general tendency to measure $c_{\rm fit}< c_{\rm V}$ for the bulk of halos. 
The right panel of Fig.~\ref{fig:compsim} shows that $\chi^2$ is positively correlated with the mass. On the basis of these results, we must conclude that {\em for a large fraction of halos, and for the most massive in particular, the NFW functional does not represent a good fit to the density profile}. As a consequence, the value of the resulting $c_{fit}$ is dependent on the radial range chosen for the fit. 


\section{Summary and conclusions}
\label{sect:conclu}
In this paper, we considered two among the cluster-scale $c-M$ relations proposed in the literature that exhibit the largest mismatch: those proposed by Prada et al. (2012) and Duffy et al. (2008). These relations differ mainly in two aspects: $i)$  the amplitude of the D08 relation is lower by $\sim 40\%$; $ii)$ the $c-M$ relation of $P12$ exhibits an upturn at the highest masses which is not present in the D08 relation and in other works. 
By means of both analytic and numerical models, we demonstrated that the mismatch can be explained by differences in the procedures for binning the halos and for measuring the concentrations.  In more details, we found that:
\begin{itemize}
\item if the $c-M$ relation is constructed using halos binned by their maximum circular velocity (P12) higher concentrations are expected at the same mass scale than for mass-selected halos (D08). This is true independently of the method followed to measure the concentrations. We found that the difference between $c_{M sel}$ and $c_{V sel}$ grows with the scatter of the concentration distribution.  For values of the scatter typically found in numerical simulations ($\sigma\lesssim 0.4$), the expected impact of the halo selection method on the $c-M$ relation is $\sim 15-20\%$ and it is mass-dependent: the highest masses are mostly affected by the binning method (see also appendix of P12). Binning can also explain, at least partially, the upturn seen in the P12 $c-M$ relation; 
\item the concentrations derived by fitting the halo density profiles with NFW models (Eq.~\ref{eq:nfw}, D08) and by means of the $V$-ratio  (Eq.~\ref{vmaxc}, P12) agree well with each other only if the NFW fit is performed over the radial range $R_{max}\leq R \leq R_{200}$, which corresponds to the scales where the $V$-ratio is evaluated. When we fitted the Multidark halos  over the same radial range used by D08, we found that the agreement between concentration measurements is good only for the halos with the smallest best-fit $\chi^2$. However, the bulk of halos shows a tendency for larger concentrations when using the P12 method. We also found that the best-fit $\chi^2$ is positively correlated with the halo mass. Typically, the most massive halos have a worse NFW-fit because they are generally far from equilibrium. This helps further to explain the upturn in the $c-M$ relation of P12.  We verified that the method to measure the concentrations impacts for another $\sim 20\%$ on the amplitude of the $c-M$ relation for halos at $z\sim 0.25$.
 \end{itemize}
On the basis of these results, we can conclude that the relations proposed by P12 and D08  can be fully reconciled if one takes into account the impact of both the halo selection and the method to measure the concentrations. An important caveat  that emerges from our analysis is that, to compare results, it is relevant to measure concentrations in a consistent manner.
This is even more crucial when we aim at comparing theoretical and observational results. Indeed, different observations probe the halo mass profiles over different radial ranges \citep{2007MNRAS.379..190C,2011A&A...526C...1E,2007MNRAS.379..209S,2012MNRAS.420.3213O,2012MNRAS.tmp.3239F}. To this issue we dedicated an entire companion paper \citep{2013arXiv1301.7476R}. 

\section*{AKNOWLEDGEMENTS}
We gratefully acknowledge  financial support from ASI through the contract EUCLID-IC phase B2/C, from INAF through the grant PRIN-INAF 2009 and from the National Science Foundation through grant AST-1210973. We thank L. Moscardini, S. Ettori, C. Giocoli, D. Lemze, J. Merten, S. Borgani and the whole CLASH team for their helpful comments and suggestions.

\bibliographystyle{/Users/maxmen/Documents/Pub/Papers/TeXMacro/mn2e}
\bibliography{/Users/maxmen/Documents/Pub/Papers/TeXMacro/master}

\begin{thebibliography}{27}
\expandafter\ifx\csname natexlab\endcsname\relax\def\natexlab#1{#1}\fi

\bibitem[{{Bhattacharya} {et~al}\mbox{.}(2011){Bhattacharya}, {Habib}, \&
  {Heitmann}}]{2011arXiv1112.5479B}
{Bhattacharya} S., {Habib} S., {Heitmann} K., 2011, ArXiv e-prints

\bibitem[{Bullock {et~al}\mbox{.}(2001)Bullock, Kolatt, Sigad, Somerville,
  Kravtsov, Klypin, Primack, \& Dekel}]{BU01.1}
Bullock J., Kolatt T., Sigad Y., Somerville R., Kravtsov A., Klypin A., Primack
  J., Dekel A., 2001, MNRAS, 321, 559

\bibitem[{{Comerford} \& {Natarajan}(2007)}]{2007MNRAS.379..190C}
{Comerford} J.~M., {Natarajan} P., 2007, \mnras, 379, 190

\bibitem[{{De Boni} {et~al}\mbox{.}(2012){De Boni}, {Ettori}, {Dolag}, \&
  {Moscardini}}]{deboni.etal.12}
{De Boni} C., {Ettori} S., {Dolag} K., {Moscardini} L., 2012, ArXiv e-prints

\bibitem[{Dolag {et~al}\mbox{.}(2004)Dolag, Bartelmann, Perrotta, Baccigalupi,
  {et~al.}}]{DO03.2}
Dolag K., Bartelmann M., Perrotta F., Baccigalupi C., {et~al.}, 2004, A\&A,
  416, 853

\bibitem[{{Duffy} {et~al}\mbox{.}(2008){Duffy}, {Schaye}, {Kay}, \& {Dalla
  Vecchia}}]{2008MNRAS.390L..64D}
{Duffy} A.~R., {Schaye} J., {Kay} S.~T., {Dalla Vecchia} C., 2008, \mnras, 390,
  L64

\bibitem[{Eke {et~al}\mbox{.}(2001)Eke, Navarro, \& Steinmetz}]{EK01.1}
Eke V., Navarro H., Steinmetz M., 2001, ApJ, 554, 114

\bibitem[{{Ettori} {et~al}\mbox{.}(2011){Ettori}, {Gastaldello}, {Leccardi},
  {Molendi}, {Rossetti}, {Buote}, \& {Meneghetti}}]{2011A&A...526C...1E}
{Ettori} S., {Gastaldello} F., {Leccardi} A., {Molendi} S., {Rossetti} M.,
  {Buote} D., {Meneghetti} M., 2011, \aap, 526, C1

\bibitem[{{Fedeli}(2012)}]{2012MNRAS.tmp.3239F}
{Fedeli} C., 2012, \mnras, 3239

\bibitem[{{Frenk} {et~al}\mbox{.}(1999){Frenk}, {White}, {Bode}, {Bond},
  {Bryan}, {Cen}, {Couchman}, {Evrard}, {Gnedin}, {Jenkins}, {Khokhlov},
  {Klypin}, {Navarro}, {Norman}, {Ostriker}, {Owen}, {Pearce}, {Pen},
  {Steinmetz}, {Thomas}, {Villumsen}, {Wadsley}, {Warren}, {Xu}, \&
  {Yepes}}]{frenk99}
{Frenk} C.~S. {et~al.}, 1999, \apj, 525, 554

\bibitem[{{Gao} {et~al}\mbox{.}(2008){Gao}, {Navarro}, {Cole}, {Frenk},
  {White}, {Springel}, {Jenkins}, \& {Neto}}]{2008MNRAS.387..536G}
{Gao} L., {Navarro} J.~F., {Cole} S., {Frenk} C.~S., {White} S.~D.~M.,
  {Springel} V., {Jenkins} A., {Neto} A.~F., 2008, \mnras, 387, 536

\bibitem[{{Klypin} {et~al}\mbox{.}(2011){Klypin}, {Trujillo-Gomez}, \&
  {Primack}}]{2011ApJ...740..102K}
{Klypin} A.~A., {Trujillo-Gomez} S., {Primack} J., 2011, \apj, 740, 102

\bibitem[{{Kravtsov} {et~al}\mbox{.}(1997){Kravtsov}, {Klypin}, \&
  {Khokhlov}}]{1997ApJS..111...73K}
{Kravtsov} A.~V., {Klypin} A.~A., {Khokhlov} A.~M., 1997, \apjs, 111, 73

\bibitem[{{Ludlow} {et~al}\mbox{.}(2012){Ludlow}, {Navarro}, {Li}, {Angulo},
  {Boylan-Kolchin}, \& {Bett}}]{2012arXiv1206.1049L}
{Ludlow} A.~D., {Navarro} J.~F., {Li} M., {Angulo} R.~E., {Boylan-Kolchin} M.,
  {Bett} P.~E., 2012, ArXiv e-prints

\bibitem[{{Macci{\`o}} {et~al}\mbox{.}(2008){Macci{\`o}}, {Dutton}, \& {van den
  Bosch}}]{2008MNRAS.391.1940M}
{Macci{\`o}} A.~V., {Dutton} A.~A., {van den Bosch} F.~C., 2008, \mnras, 391,
  1940

\bibitem[{{Mu{\~n}oz-Cuartas} {et~al}\mbox{.}(2011){Mu{\~n}oz-Cuartas},
  {Macci{\`o}}, {Gottl{\"o}ber}, \& {Dutton}}]{2011MNRAS.411..584M}
{Mu{\~n}oz-Cuartas} J.~C., {Macci{\`o}} A.~V., {Gottl{\"o}ber} S., {Dutton}
  A.~A., 2011, \mnras, 411, 584

\bibitem[{Navarro {et~al}\mbox{.}(1996)Navarro, Frenk, \& White}]{NA96.1}
Navarro J., Frenk C., White S., 1996, ApJ, 462, 563

\bibitem[{Navarro {et~al}\mbox{.}(1997)Navarro, Frenk, \& White}]{NA97.1}
Navarro J., Frenk C., White S., 1997, ApJ, 490, 493

\bibitem[{{Oguri} {et~al}\mbox{.}(2012){Oguri}, {Bayliss}, {Dahle}, {Sharon},
  {Gladders}, {Natarajan}, {Hennawi}, \& {Koester}}]{2012MNRAS.420.3213O}
{Oguri} M., {Bayliss} M.~B., {Dahle} H., {Sharon} K., {Gladders} M.~D.,
  {Natarajan} P., {Hennawi} J.~F., {Koester} B.~P., 2012, \mnras, 420, 3213

\bibitem[{{Prada} {et~al}\mbox{.}(2012){Prada}, {Klypin}, {Cuesta},
  {Betancort-Rijo}, \& {Primack}}]{2012MNRAS.tmp.3206P}
{Prada} F., {Klypin} A.~A., {Cuesta} A.~J., {Betancort-Rijo} J.~E., {Primack}
  J., 2012, \mnras, 3206

\bibitem[{{Rasia} {et~al}\mbox{.}(2013){Rasia}, {Borgani}, {Ettori},
  {Mazzotta}, \& {Meneghetti}}]{2013arXiv1301.7476R}
{Rasia} E., {Borgani} S., {Ettori} S., {Mazzotta} P., {Meneghetti} M., 2013,
  ArXiv e-prints

\bibitem[{{Schmidt} \& {Allen}(2007)}]{2007MNRAS.379..209S}
{Schmidt} R.~W., {Allen} S.~W., 2007, \mnras, 379, 209

\bibitem[{Sheth \& Tormen(2002)}]{SH02.1}
Sheth R., Tormen G., 2002, MNRAS, 329, 61

\bibitem[{Springel {et~al}\mbox{.}(2001)Springel, Yoshida, \& White}]{SP01.1}
Springel V., Yoshida N., White S., 2001, New Astronomy, 6, 79

\bibitem[{{Wechsler} {et~al}\mbox{.}(2002){Wechsler}, {Bullock}, {Primack},
  {Kravtsov}, \& {Dekel}}]{2002ApJ...568...52W}
{Wechsler} R.~H., {Bullock} J.~S., {Primack} J.~R., {Kravtsov} A.~V., {Dekel}
  A., 2002, \apj, 568, 52

\bibitem[{Zhao {et~al}\mbox{.}(2003)Zhao, Jing, Mo, \& B\"orner}]{ZH03.1}
Zhao D., Jing Y., Mo H., B\"orner G., 2003, ApJ, 597, L9

\bibitem[{{Zhao} {et~al}\mbox{.}(2003){Zhao}, {Jing}, {Mo}, \&
  {B{\"o}rner}}]{zhao2003}
{Zhao} D.~H., {Jing} Y.~P., {Mo} H.~J., {B{\"o}rner} G., 2003, \apjl, 597, L9

\end{thebibliography}
\label{lastpage}
\end{document}